# FLASH: Randomized Algorithms Accelerated over CPU-GPU for Ultra-High Dimensional Similarity Search

Yiqiu Wang
Rice University
Houston, Texas
yiqiu.wang@rice.edu

Anshumali Shrivastava
Rice University
Houston, Texas
anshumali@rice.edu

Jonathan Wang
Rice University
Houston, Texas
jw96@rice.edu

Junghee Ryu
Rice University
Houston, Texas
jr51@rice.edu

## ABSTRACT

We present FLASH (**F**ast **L**SH **A**lgorithm for **S**imilarity search accelerated with **H**PC), a similarity search system for ultra-high dimensional datasets on a single machine, that does not require similarity computations and is tailored for high-performance computing platforms. By leveraging a LSH style randomized indexing procedure and combining it with several principled techniques, such as reservoir sampling, recent advances in one-pass minwise hashing, and count based estimations, we reduce the computational and parallelization costs of similarity search, while retaining sound theoretical guarantees.

We evaluate FLASH on several real, high-dimensional datasets from different domains, including text, malicious URL, click-through prediction, social networks, etc. Our experiments shed new light on the difficulties associated with datasets having several million dimensions. Current state-of-the-art implementations either fail on the presented scale or are orders of magnitude slower than FLASH. FLASH is capable of computing an approximate k-NN graph, from scratch, over the full webspam dataset (1.3 billion nonzeros) in less than 10 seconds. Computing a full k-NN graph in less than 10 seconds on the webspam dataset, using brute-force ($n^2D$), will require at least 20 teraflops. We provide CPU and GPU implementations of FLASH for replicability of our results[1].

[1]https://github.com/RUSH-LAB/Flash

## 1 INTRODUCTION

Similarity search, or k-nearest-neighbor search (k-NNS), is one of the most frequent operations in large scale data processing systems. Given a query object $q$, with feature representation $q \in \mathbb{R}^D$, the goal of similarity search is to find, from a collection $C$ of $N$ data instances, an object $x$ (or set of objects) most similar to the given query. The notions of similarity are based on some popular Euclidian type measures such as cosine similarity [11] or Jaccard similarity [8].

**k-NNS over Ultra-high Dimensional and Sparse Datasets:** Recommendation systems naturally deal with ultra-high dimensional and sparse features as they usually consist of categorical combinations. Even the general user-item matrix representation leads to ultra-sparse and ultra-high dimensional representation. Neighborhood models [15] are popular in recommendation systems where the first prerequisite is to find a set of near-neighbor for every item. Social networks are another natural ground for ultra-high dimensional and extremely sparse representations. In social networks, we represent the friendship relations between users as graphs. Given $d$ users, we describe each user as a $d$ ultra-high dimensional and very sparse vector, whose non-zero entries correspond to edges. By representing each user as a column, we construct matrix $A$ of dimension $d \times d$. Finding similar entries of such a user representation is one of the first operations required for a variety of tasks including link prediction [26], personalization [14], and other social network mining tasks [47]. Other popular applications where k-NNS over ultra-high dimensional and sparse dataset is common include click-through predictions [32] and plagiarism detection [7].

The naive way to perform k-NNS is to compute the exact distance (or similarity) between the query and all data points, followed by ranking. The naive approach suffers from time complexity of $O(N \cdot D)$ per query, which is known to be computationally prohibitive when $N$ is huge and querying is a frequent operation [46]. If we treat each object in collection $C$ as the query, then the result of $N$ k-NNS query leads to what is also known as a $k-NN$ graph which has a computation complexity of $N^2 \cdot D$, a significantly expensive operation for massive datasets.

**Approximations Suffice In Practice:** If we allow approximations, then it is algorithmically possible to obtain efficient solutions.

Fortunately, in practice, it is sufficient to solve the similarity search problem approximately, at the benefit of reduced latency, which is a critical factor in many applications. Not surprisingly, efficient and approximate near-neighbor search has been studied extensively in the past. Due to its ubiquitous nature in several different settings, approximate k-NNS is still an active area of research in the databases and data mining community.

**Existing Approaches and Shortcomings:** There are several strategies for approximate near-neighbor search. Earlier methods focused on deterministic space partitioning, such as kd-trees [9], which were suitable for low dimensional datasets [46]. These methodologies suffer from the curse of dimensionality leading to poor performance, reducing the search to near-linear scan on high-dimensional datasets. Randomized indexing approaches based on locality sensitive hashing (LSH) [19] showed significant promise in dealing with the curse of high-dimensionality.

Hash tables based on LSH are known to have skewed bucket sizes, where a large number of buckets in hash tables are near-empty while some other buckets are quite heavy. This skewness hurts the efficiency of the algorithm, and furthermore, it leads to uneven load balancing making parallelization less useful. Also, LSH is known to require a significant amount of memory for storing multiple hash tables [34]. Nevertheless, LSH is still widely adopted because hash tables are simple and easy to maintain.

Two notable strategies which have gained popularity recently are: 1) Product Quantization (PQ) [20] and 2) Random Proximity Graph Method [30]. PQ partitions the dimensions into small subgroups and then pre-computes the set of neighbors on each partition as a lookup table. Neighbors from these sub-groups comprise a suitable candidate set. In [21], authors used product quantization over four general purpose graphics processing units (GPU) to show impressive performance on billion scale image datasets, with relatively small (128) dimensions.

However, PQ, similar to other deterministic space partitioning ideas, suffers from the curse of dimensionality. For ultra-high dimensional (several million) and significantly sparse vectors, such as those of interest in this work, PQ fails. At several million dimensions, candidates generated by looking at only a small set of dimensions are unlikely to be useful, since any small subset is likely to be all zeros due to sparsity. Besides, the memory footprint of PQ scales linearly with the dimensionality. The linear scaling is because every disjoint subset of dimensions requires its own lookup tables. Not surprisingly, our experiments show that the recently proposed, PQ based, FAISS system which is optimized for GPUs runs out of memory on the datasets of interest to the paper. See section 4.4.4 for details.

Currently the state-of-the-art approximate near-neighbor technique is based on the navigable small world (NSW) graph concept [30]. The idea is borrowed from connectivity in social networks, which exhibits random short and long links. The NSW algorithm constructs a similar structure based on a variant of greedy branch-and-bound type navigating algorithms [31] that iterates through pre-computed proximity graphs with polylogarithmic complexity. A more advanced version, the hierarchical NSW (HNSW) algorithm [31] improves the performance and allows a logarithmic complexity search by using a hierarchical set of layers.

Existing methods, including NSW and HNSW, rely on constructing an expensive and memory heavy data structure that helps in quickly pruning down candidates (See Appendix E). There is usually a tradeoff between memory and query time computation, and one can be sacrificed for the other. However, the index building time and memory are equally important, because they are related to the cost of managing the data structure (update time). Unfortunately, for most data structures, such as kd-trees, insertion is hard to parallelize. As a result, there has been an increased emphasis on the total time for full k-NN graph construction time from scratch which includes the data structure construction time, rather than just the query time. It is also known that the same complexity of computing approximate k-NN graphs is the fundamental bottleneck in large-scale deep learning [41]. Furthermore, with increased dimensionality, pruning becomes less effective, which results in a lot of similarity computations to eliminate candidates reliably. Overall these two requirements become the major bottleneck both in maintaining the data structure as well as querying.

**High-Performance Computing Platforms such as GPUs:** Parallelization is one of the essential components of modern big-data processing systems. Theoretical efficiency is not sufficient for practicality if the algorithm is not amenable to parallel speedups. GPU platforms are gaining popularity because they are cheap, and now an integral part of any data processing system. As a result, there is an increased interest in implementations of k-NN graph on GPUs [21]. It is beyond doubt that being able to utilize these platforms efficiently makes a huge difference in practice and adoption.

A notable GPU based k-NN graph implementation is FAISS [21], where the authors leverage four GPUs, instead of one, and show impressive performance over 128-dimensional image features, using product quantization (PQ). As noted before, the PQ technique is not suitable for ultra-high dimensional data, which is the focus of this paper. Our experiments indicate that the FAISS library runs out of memory on several million dimensional datasets.

GPUs are mostly memory constrained. For example, one of the best units available, the Nvidia Tesla P100 is limited to 16 GB and cannot even store some of the large scale datasets. Thus, any methodology relying on similarity computation will have to bring the data to the GPU device memory for similarity computations with the query. Unfortunately, most methods, including HNSW, require similarity computation between the query and several sets of points to report top-k neighbors, which incurs unavoidable data movements to GPU memory. FAISS avoids such distance computations by using PQ based estimation, but it is not suitable for million dimensional datasets as mentioned.

**Even Dimensionality Reduction is Slow:** Since we allow approximations, a reasonable strategy is first to perform dimensionality reduction and then apply any k-NNS method over low-dimensional data. However, dimensionality reduction for ultra-high dimensional datasets is a costly operation. Even utilizing smart random projections in parallel can be significantly demanding. As an illustration, FLASH is capable of computing an approximate k-NN graph 42 times faster than the mere computation of a 100-dimensional sparse random projection of the data in parallel on the same machine. In addition, for millions of dimensions, the size of random numbers is often larger than the data, if the data is very sparse.

Overall, an efficient, low-memory, and scalable approximate nearest-neighbor search system for ultra-high dimensional datasets requires balancing several different aspects. We believe this paper provides such a system.

**Our Focus:** Our focus is on both approximate k-NN graph computation and approximate k-NNS querying over ultra-high dimensional datasets, which are commonly seen in practice. We limit ourselves to single machine implementations which exploit parallelism available in the form of multi-cores and/or a GPU.

We note that there are several efficient implementations and modifications of LSH [42] on a distributed and streaming setting, which is not the focus of this paper. We further stress that most LSH implementations [33, 44] use random projection based LSH, which we argue is significantly slower for our requirements.

## 1.1 Our Contributions:

We propose an LSH algorithm for similarity search tailored for high-performance platforms, which does not require any similarity computations. Being similarity computation free, FLASH does not need to store the data features and hence is significantly more memory efficient.

FLASH is a combination of several novel modifications to the LSH based similarity search algorithm, which is carefully tailored to balance computational cost, parallelizability, and accuracy. Our unique choices of hash function combined with reservoir sampling and collision-based ranking provably eliminates some of the frequently encountered problems associated with LSH, such as variable sized and growing buckets, which could be of independent interest in itself. Due to randomized insertions and the use of online procedures only, our process is massively data parallel with a very low chance of conflict between processors. We naively parallelize the entire algorithm using OpenMP as a first step. We also implemented an OpenCL version with a focus on k-selection, the main bottleneck of the KNN graph construction and achieved around a 1.5-3.5x speedup, on a 56 threaded machine, by additionally utilizing a GPU (NVIDIA Tesla P100).

We provide substantial empirical evidence on four real ultra-high dimensional datasets coming from email documents, URLs, click-through predictions, and social network graphs. Our experiments indicate improvements with FLASH over state-of-the-art alternatives.

## 2 BACKGROUND

### 2.1 Minwise Hashing and Locality Sensitive Hashing (LSH) Algorithm

The oldest and most famous locality sensitive hashing (LSH) scheme is minwise hashing [8] which works over binary vectors. The original minwise hashing computations require a random hash function $\pi : N \rightarrow N$ (or permutation), from integers to integers. Corresponding to this hash function $\pi$, the minwise hash $h_\pi$ for any $x \in {0, 1}^D$ is given by:

$$h_\pi(x) = \min_{i \text{ s.t. } x_i \neq 0} \pi(i) \quad (1)$$

Under the randomization of $\pi$, the probability that the minwise hash values of two different $x$ and $y$ agree is precisely the Jaccard similarity between $x$ and $y$. Formally,

$$Pr(h_\pi(x) = h_\pi(y)) = \frac{x \cdot y}{|x| + |y| - x \cdot y}, \quad (2)$$

where $|x|$ is the number of non-zeros elements in $x$. The quantity $\frac{x \cdot y}{|x|+|y|-x \cdot y}$ is the famous Jaccard Similarity.

It has been recently shown that minwise hashing is both theoretically and empirically preferred hash function over signed random projection [38] even for cosine similarity measure. The cosine similarity for binary vectors is given by

$$C(x, y) = \frac{x \cdot y}{\sqrt{|x||y|}} \quad (3)$$

### 2.2 (K,L)-parameterized LSH Algorithm

Since our algorithm builds on the classical $(K, L)$-parameterized LSH Algorithm, we describe the process briefly. For more details please refer to [5]. The Algorithm requires $L$ random meta-hash functions, $H_i$ $i = \{1, 2, ..., L\}$. Each of this meta hash function $H_i$ is formed from $K$ different LSH hash functions. Formally, each meta-hash function can be thought of as a K-tuple value, $H_i = \{h_{i,1}, h_{i,2}, ..., h_{i,K}\}$, where each $h_{i,j}$ is an LSH hash function, such as minwise hashing. Overall, we need a total of $K \times L$ LSH hash signatures of the data.

With these $L$ meta-hash functions, the Algorithm works in two phases: 1) Adding or Hash table insertion phase and 2) Querying or search phase. The querying phase can be further divided into two sub-phases: a) Candidate Generation and b) Top-k selection.

- **Adding Phase:** We create $L$ different hash tables, where every hash key points to a bucket of elements. For every element $x$ in the collection $C$, we insert $x$ (identifiers only) in the bucket at location $H_i(x)$ in table $i = \{1, 2, ..., L\}$. To assign K-tuples $H_i$ to a location, we use some universal random mapping function to the desired address range. See [5] for details.
- **Query Phase:** Given a query $q$ whose neighbors we are interested in:
  - **Candidate Generate Phase:** From table $i$, get all elements in the bucket addressed by $H_i(q)$, where $i = \{1, 2, ..., L\}$. Take union all the $L$ buckets obtained from $L$ hash tables.
  - **Top-K Selection:** From the selected candidates, report the top-k candidates based on similarity with $q$.

### 2.3 Densified One Permutation Hashing (DOPH)

Computing several minwise hashes of data is a very costly operation [24]. Fortunately, recent lines of work [38] on Densified One Permutation Hashing (DOPH) have shown that it is possible to compute several hundreds of even thousands, hashes of the data vector in one pass with nearly identical properties as minwise hashes. We will use the most recent variant [36] as our datasets are very sparse. Our experiment shows that computing DOPH is disruptively faster compared to all other hashing schemes. Our hashing mechanism throughout the paper will be DOPH.

We note that DOPH is a significant advancement critical for FLASH. On ultra-high dimensional datasets, such as *webspam* with

16 million dimensions and 3700 nonzeros, the cost of random projection can be more than 400x slower than DOPH. See section 4.4.6 for direct comparisons. DOPH only requires 4 random numbers to generate all the hashes in one pass [36]. On the other hand, with random projections, over 16 million dimensions, the cost of storing and accessing projection matrix is a huge burden. Furthermore, 100 random projections require to loop over the data vector 100 times, even if we use the fast variant of Achlioptas [3] which avoids multiplications and uses sparsity to reduce computation further.

As a result, FLASH can compute full k-NN graph computation significantly faster than calculating 100 random projections in parallel using 56 threads. We reiterate that LSH based on random projections requires around thousands or more projections for the dataset used in this paper. Thus, using other LSH approaches would not lead to the performance demonstrated in this work.

## 2.4 Reservoir Sampling

Vitter's reservoir sampling algorithm [43] processes a stream of $m$ numbers and can generate $R$ uniform samples of the given stream by only using an array of size $R$. The process is outlined in Algorithm 1. The algorithm only needs one pass over the stream.

---

**Algorithm 1** Reservoir Sampling

1: **procedure** ReservoirSampling($Reservoir[0\ldots R-1]$, $Stream[0\ldots m-1]$)
2:     **for** $i = [0, R-1]$ **do**
3:         $Reservoir[i] := S[i]$
4:     **end for**
5:     **for** $i = [R, m-1]$ **do**
6:         $j := \text{Random}([0, i])$
7:         **if** $j \leq R$ **then**
8:             $Reservoir[j] := S[i]$
9:         **end if**
10:    **end for**
11: **end procedure**

---

## 3 PROPOSED ALGORITHM

### 3.1 Issues with LSH Algorithm

We first focus on several issues, which limit the efficiency and scalability of LSH algorithms for ultra-high dimensional and sparse data sets:

(1) **Hash Computation Cost:** The ($K$, $L$) parameterized LSH Algorithm requires $K \times L$ hash calculations of the data, which are usually into hundreds or more. With traditional LSH it will need hundreds or more passes over the data, a prohibitively expensive operation. Hashing cost is a known computational bottleneck with LSH [28].

(2) **Skewed Buckets:** Since the LSH bucket assignment is random and data dependent. After hashing, the bucket sizes are heavily skewed in practice. The bucket sizes cannot be inferred in advance as the process is dynamic. Skewed buckets have two issues: 1) We need to rely on some dynamically increasing data structure. Such a data structure has additional resizing overheads. 2) If we perform bucket aggregation in parallel, then skewed buckets are hard to parallelize due to unequal distribution of work.

(3) **Similarity Computations and Data Storage:** LSH algorithms use candidate similarity calculations to report top-k neighbors. Thus, we need to store the complete datasets which should be brought into main memory whenever needed. With GPUs this is more critical as most of the data will not fit the GPU memory and will require switching.

(4) **GPUs are not suitable for Sparse Datasets:** A known issue with sparse datasets is that they have poor performance on GPUs as memory coalescing is not readily available. For ultra-high dimensional sparse datasets, dense representation of vectors will blow up the memory into terabytes or beyond. As an instance, one of our datasets, the webspam dataset, has 16 million dimensions. It requires around 10GB in binary indexing format. Converting it into dense format will require more than a terabyte of space. Thus, we are forced to work with sparse indexing formats. Sparse operations are not particularly impressive over GPUs.

### 3.2 Our Proposed Fix

We first focus on the proposed modifications made to the LSH algorithm to address the above-mentioned efficiency and scalability issues. We will then discuss their theoretical justifications in section 3.3. We made the following specific algorithmic changes:

**1. Densified One Permutation Hashes (DOPH):** We use the recent advances in densified one permutation hashing (DOPH) [38] which computes hundreds of minwise hashes in one pass over the data vector. DOPH is ideally suited for ultra-high dimensional and sparse dataset. LSH based on random projections are other alternatives. There are intelligent strategies on making hash computations faster such as ones using Fast Walsh—Hadamard transformations [4]. However, with several million incredibly sparse dimensions, Fast Walsh—Hadamard is quite costly as it makes the data dense (requires dense matrix multiplication of the order of dimensions). Sparse projections, although appealing, are slow and have significant memory overheads for storing random numbers.

Overall, with DOPH, we convert sparse data into $K \times L$ hash signatures in a single pass using a simple hash function as shown in [36]. These fixed sized representations can be readily used in specialized devices such as GPUs. As we will show, in the querying phase, our algorithm never needs the indexed data and can work with only hash tables, which are significantly smaller than the data size. In section 3.3, we give yet another reason for using DOPH.

**2. Fixed Sized Reservoir Sampling of Buckets:** The skewness of the buckets is dependent on the data distribution, which we cannot know in advance. A simple but principled modification is instead of using dynamically growing buckets, we only keep a fixed size reservoir (simple arrays), and use online reservoir sampling to obtain a uniform sample of the bucket. We show in Section 3.3 that this random sample of the bucket is sufficient, and even for a small reservoir size, the procedure does not affect the theoretical guarantees on LSH algorithm in any way.

With a small fixed sized array, the buckets are never too crowded, and they also provide ideal load balancing of threads during parallel bucket aggregation. The advantages come without any insertion overhead. We only need a couple of random number generations per insertion. Also, the process has strong theoretical guarantees.

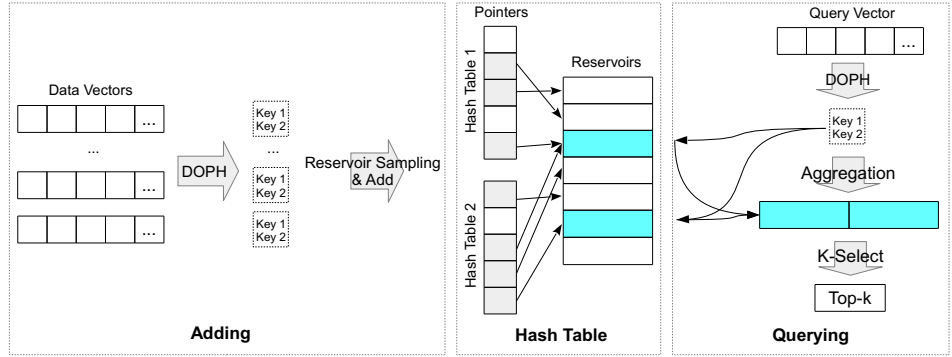

Figure 1: Algorithm Overview: A illustration with $L = 2$ Hash Tables.

**3. Count based $k$-selection:** We make an observation that with $L$ different hash tables, data points that appear more frequently in the aggregated reservoirs are more likely to be similar to the query data point. This observation allows us to estimate the actual ranking unbiasedly. We count the frequency of occurrence of each data point in the aggregated reservoirs. Based on this count, we report the $k$ most frequent as nearest neighbors. We call this process *count based k-selection*. We note that estimating similarity instead of calculating it is not new [16, 37]. Our collision counting approach can be efficiently utilized on GPUs without any additional memory overheads; see section 3.4.1 for details.

**4. Reservoir Sharing across Hash Tables:** In LSH, most buckets are quite sparse (near empty) due to a large number of buckets. With reservoir sampling, we eliminate the problem of overcrowded buckets. However, with fixed bucket size, near empty buckets create memory overhead. For better utilization of space, we allow random reservoir (or bucket) sharing across hash tables. In particular, we allocate a small local pool of reservoirs for each hash table. If the allocation is exhausted, then reservoirs from the global pool are assigned randomly, i.e., every hash table picks one of the reservoirs as its own, as shown in Figure 3.

Two hash tables sharing the same bucket only hurts significantly if both buckets are heavy. However, heavy buckets are rare, and collision of two heavy buckets is exceedingly rare. Reservoir sharing, therefore, has very small impact on the search quality.

It should be noted that most of our memory requirements are already low, but reservoir sharing gives us another 5x factor of improvement without apparent loss in accuracy. Due to randomness, the procedure does not hurt data parallelism.

Overall, the complete process is summarized in Algorithm 2 and Algorithm 3. Also, see Figure 1 for an illustration of the entire process. We defer the description of reservoir sharing to section 3.5 for ease of understanding. Just like the $(K, L)$-parameterized LSH algorithm, our algorithm also consists of an *Adding phase* and a *Querying phase* with the above mentioned algorithmic modification. The *Adding phase* computes hashes for input data points and stores their identifiers in the hash tables. With every addition of the data, we can discard the data completely and only work with hash table addresses and identifiers. In the *Querying phase*, reservoirs are aggregated based on the hashes of the query. *K-selection* is then performed on the aggregated reservoirs which only uses the frequency of stored identifiers to report the top-$k$ neighbors. Thus, for answering any query, we only need to know the $L$ hash addresses instead of storing the raw data.

---

**Algorithm 2** The Adding Phase

1: **procedure** ADDING-PHASE
2:   **for** each *DataPoint* **do**
3:     $AllHashes := \text{DOPH}(DataPoint)$
4:     **for** each $Table_i$ **do**
5:       $Key_i := \text{MapKHashesToAddress}$
6:       ADD($DataPoint_i, Table_i, Key_i$)
7:     **end for**
8:   **end for**
9: **end procedure**

10: **function** ADD($DataPoint, Table_i, Key$)
11:   **if** $Table_I[Key]$ Empty **then**
12:     $Table_I[Key] = \text{AllocateReservoir}$
13:     ReservoirCounter $=0$
14:   **end if**
15:   $Rand := \text{RANDOM}([0, ReservoirCounter])$
16:   **if** $Rand < R$ **then**
17:     $Reservoir[Rand] = DataPoint$
18:   **end if**
19:   ReservoirCounter++
20: **end function**

---

### 3.3 Theoretical Justification

We now argue why our approach has solid theoretical justification. We briefly review a few definitions and the sub-linearity results associated with classical LSH. Proofs are deferred to the appendix for better readability.

DEFINITION 1. *(c-Approximate Near Neighbor or c-NN). Consider a set of n points, denoted by C, in a D-dimensional space $\mathbb{R}^D$, and parameters $S_0 > 0$, $\delta > 0$. The task is to construct a data structure*

**Algorithm 3** The Querying Phase

```
1: procedure QUERYING-PHASE
2:     for each QueryPoint do
3:         AllHashes := DOPH(QueryPoint)
4:         Initialize A
5:         for each Table_i do
6:             Append A with Table_t[Key]
7:         end for
8:         Output_i := KSELECT(A)
9:     end for
10: end procedure

11: function KSELECT(A)
12:     SORTINPLACE(A)
13:     KVPair = COUNTFREQUENCY(A)
14:     SORTBYVALUEINPLACE(KVPair)
15:     return KVPair[0:TopK]
16: end function

17: function COUNTFREQUENCY(A)
18:     INITIALIZE KVPair
19:     for each Key in A do
20:         if Key_i == Key_{i-1} then
21:             KVPair[Key_i]++
22:         end if
23:     end for
24:     return KVPair
25: end function
```

which, given any query point $q$, if there exist an $S_0$-near neighbor of $q$ in $P$, it reports some $cS_0$-near neighbor of $q$ in $C$ with probability $1 - \delta$.

The usual notion of $c$-NN is for distance. Since we deal with similarities, we define $S_0$-near neighbor of point $q$ as a point $p$ with $Sim(q, p) \geq S_0$, where $Sim$ is the similarity function of interest such as cosine similarity.

DEFINITION 2. *(Locality Sensitive Hashing (LSH)) A family $\mathcal{H}$ is called $(S_0, cS_0, p_1, p_2)$-sensitive if for any two point $x, y \in \mathbb{R}^D$ and $h$ chosen uniformly from $\mathcal{H}$ satisfies the following:*
- *if $Sim(x, y) \geq R_0$ then $Pr_{\mathcal{H}}(h(x) = h(y)) \geq p_1$*
- *if $Sim(x, y) \leq cR_0$ then $Pr_{\mathcal{H}}(h(x) = h(y)) \leq p_2$*

For approximate nearest neighbor search typically, $p_1 > p_2$, and $c < 1$ is needed. Note, $c < 1$, as we are defining neighbors in terms of similarity.

FACT 1. *Given a family of $(S_0, cS_0, p_1, p_2)$-sensitive hash functions, one can construct a data structure for $c$-NN with $O(n^\rho \log_{1/p_2} n \log \frac{1}{\delta})$ query time and space $O(n^{1+\rho} \log \frac{1}{\delta})$, where $\rho = \frac{\log 1/p_1}{\log 1/p_2} < 1$.*

Most of the popular LSH algorithms satisfy a stronger condition known as monotonicity.

DEFINITION 3. *Monotonic LSH. We will call an LSH family $\mathcal{H}$ monotonic with respect to the similarity $Sim$ iff $Pr_{\mathcal{H}}(h(x) = h(y)) \geq Pr_{\mathcal{H}}(h(u) = h(v)) \iff Sim(x, y) \geq Sim(u, v)$.*

*3.3.1 Guarantees with Reservoir Sampling.* As mentioned reservoir sampling produces a fixed size random sample of all elements inserted at a hash location, and therefore, it still has most of the probabilistic guarantees intact. We can easily redo the proofs, if the underlying LSH is DOPH, by tweaking the margins in failure probability, to account for this sampling.

In particular, we show that reservoir sampling of LSH buckets, with DOPH (or minwise hashing) as LSH, does not affect the worst case asymptotic guarantees, for reservoir of size $R$ of small constant (like 5). We only need a very mild additional assumption to take care of the correlations.

**Assumption:** Given a query $q$, and any point $x$ with $Sim(q, x) \geq S_0$. We need an assumption that for any $y$ with $Sim(q, y) \leq cS_0$, we assume $Pr(h(q) = h(y)|h(q) = h(x)) \leq Pr(h(q) = h(y))$, i.e. the conditional probability is less than or equal to the unconditional probability. Note that if any $x$ and $y$ are independent of each other then this assumption is always true. We argue in Appendix A, why this assumption is almost always valid with minwise hashing.

THEOREM 1. *Under the assumption above, the LSH algorithm with reservoir sampling on the hash buckets and DOPH (or minwise hashing) as the LSH function using reservoir size satisfying $R > constant$ solves the $c$-NN instance with $O(n^\rho \log_{1/(cS_0)} n \log \frac{1}{\delta})$ query time and space $O(n^{1+\rho} \log \frac{1}{\delta})$, where $\rho = \frac{\log 1/S_0}{\log 1/(cS_0)} < 1$*

Even with $n$ around 100 million, a fixed small reservoir of size is sufficient. Our result solves the problem of having variable and large buckets, a common practical complaint with LSH. To the best of our knowledge, there is no prior work that formally addresses this problem and shows that small bucket sizes are sufficient for the same asymptotic guarantees. Small-sized buckets make a great difference from systems perspective.

*3.3.2 Count based $k$-selection.* We showed that reservoir sampling does not change any guarantees with LSH. In this section, we argue why adding count based-$k$ section is theoretically sound. We show that our overall procedure ensures an important invariant – points similar to the query have a higher probability of being in the top-$k$ than less similar ones.

Given the query $q$, define $P_{q,x}$ as the probability that $x$ is reported in top-$k$, by $(K, L)$-parameterized LSH algorithm, with monotonic LSH family, and with reservoir sampling combined with our count based $k$ selection. Let the notions of similarity associated with LSH be denoted by a two argument function $Sim(., .)$. We have the following theorem:

THEOREM 2. *For any $x, y \in C$ and for all choices of $K, L$, we have*

$$P_{q,x} \geq P_{q,y} \iff Sim(q, x) \geq Sim(q, y) \tag{4}$$

The result is also true for $L = 1$, in the individual hash tables.

For a given query $q$, let $CP(q, x)$ denote the probability of finding $x$ and $q$ in the same bucket in a given hash table (collision). Note that since all hash tables are independent, the subscript of the hash table is immaterial. From Theorem 2, we know that for any $x, y$ with $Sim(q, x) \geq Sim(q, y)$ we have $CP(q, x) \geq CP(q, y)$. Counting the occurrences of $x$, out of $L$ hash tables, in the bucket of query $q$ is, therefore, a binomial estimator of $L \times CP(q, x)$. The estimator is unbiased with variance $\frac{CP(q,x)(1-CP(q,x))}{L}$. Thus a ranking over these counts is a ranking over the estimators. Due to monotonicity, the ranking concerning $CP(q, .)$ is same as the desired ranking with respect to $Sim(q, .)$ (or the actual distance). Thus, with large enough $L$, the rankings will statistically converge to the true rankings.

The use of LSH for efficient adaptive sampling, which came to light very recently starting with [41] and followed by [10, 12, 13, 27, 40], has shown huge promise in unbiased estimation and machine learning pipelines. With theorem 2, FLASH can naturally replace LSH implementations for efficient and adaptive sampling.

**Another note on the importance of DOPH:** Since we are estimating using a small number of indices which should be small for performance, we require very accurate estimators. Estimation is another argument why DOPH is uniquely suited for FLASH. Minwise hashes are known to have sharper estimation properties [37] compared to other LSH. Minwise hashing has an extensive range space, taking $D$ (dimensions) different values. In contrast, signed random projection only produces 1 bit.

Consider the address space for 15-20 bits for hash tables. In order to obtain 15-20 bits, with high-dimensional datasets, DOPH can work with values of $K$ as small as 2-4 or even 1. However, signed random projections will require at least 15-20 hash bits, making collision in hash tables exceedingly rare. Therefore signed random projections will require large values of $L$ as most buckets will be empty. More formally, The probability of collision in hash tables gets exponentiated by $K$ [17]. The variance of the $k$-selection estimator is $\frac{1}{L}CP(q, x)^K(1 - CP(q, x)^K)$ which is poor (large) for small values of $CP(q, x)^K$. For a given similarity, signed random projection has lower values of $CP(q, x)$ and at the same time requires large $K$, compared to DOPH. As a result, the variance of ranking using $k$-selection will be significantly poor with signed random projections requiring a very large value of $L$.

## 3.4 Implementation and Parallelization

In similarity search, the query operation is trivially data parallel across multiple queries. However, with most methodology, insertion is not data parallel. Data insertion is dependent on the global state of the indices which changes with every addition. Data parallelism is crucial in order to fully utilize the massive parallelism available on modern devices which include multi-core GPUs. This is one of the unique and key qualities of LSH, which when combined with reservoir sampling to enforce load balancing, puts our implementation on top in head-to-head comparisons with other state-of-the-art packages. With LSH, the insertions are independent with high probability. Inserting $x$ and $y$ leads to a write conflict if and only if they collide in some hash table. But due to randomization, this probability is extremely low.

We provide a multi-core CPU, a full GPU and a hybrid CPU-GPU implementation of FLASH. The multi-core version exploits data-parallelism and each different thread deals with separate data instances during both addition and querying. Due to the simplicity of the procedure and load balancing of fixed-sized buckets, our multi-core implementation itself is significantly faster than the multi-core version of NMSLIB-HNSW which is the current state-of-art approximate k-NN search implementation. See Section 4.4.2 for a head-to-head comparison.

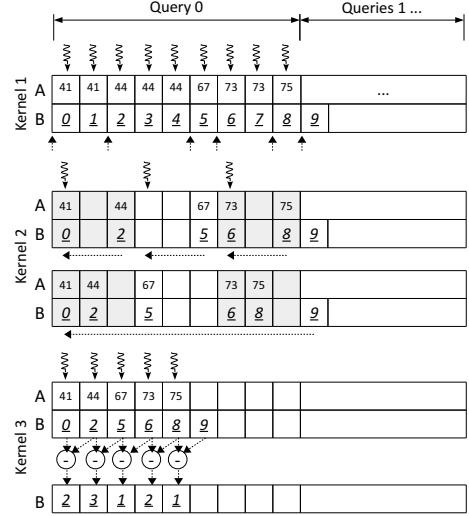

**Figure 2:** *CountReduce* on the GPU. Before kernel 1, array A is already sorted segment-wise. Kernel 3 is followed by another round of segmented sort to obtain the candidates with top counts.

In the full GPU version, the hash tables are entirely stored in the GPU on-device memory. Data instances are passed once to the GPU device during the index building phase, and once more during the querying phase. Since most of the processing is done locally on the device, this scheme provides the highest throughput given the very high memory bandwidth and computing power available on the device. The full GPU FLASH achieves a further 1.5-3.5x speed up in KNN-Graph construction on a single GPU in comparison to 56 threaded CPUs.

The CPU-GPU hybrid version of FLASH places most of the data structures in the main memory, leaving only the compute-heavy operations (the $k$-selection) to the GPUs. It comes into play when dealing with datasets too large for the GPU memory.

*3.4.1 GPU Implementation of Count Based $k$-selection.* The $k$ selection is bucket aggregation followed by sorting to report the top-$k$. Since FLASH only requires hash tables for the complete process, once they are created by CPUs in the main memory, only small chunks of the data structure need to be transferred to the GPU for k-selection processing.

Our count-based $k$-selection has input $[A_0 \ldots A_{m-1}]$, where $A_i$ are fixed-sized arrays $i$ corresponds to queries. $A_i$ is fixed size because it is a concatenation of $L$ fixed sized reservoirs. We aim to report the $k$ most frequent candidates of each $A_i$ for $i = 0 \ldots m - 1$. The algorithm can be decomposed into a counting step and a ranking step. The counting step counts the frequency of each candidate in $A_i$. The ranking step extracts the top $k$ candidates with highest counts from $A_i$.

Common approaches for the counting step include using hash maps or sorting followed by a linear pass. The former is not suited for the GPU because of potential numerous random global memory accesses while the local memory usually does not fit the hash maps;

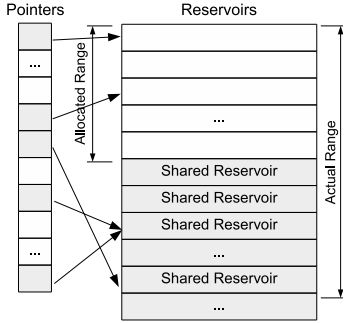

**Figure 3: Illustration of Reservoir sharing: Reservoirs are shared randomly for better space utilization.**

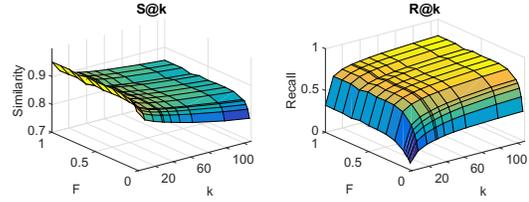

**Figure 4: Effect of varying $F$ on the search quality for dataset *url*. K=4, L=128, R=32.**

the later is naively parallelizable on the GPU, but not optimal in the linear pass step.

Instead, we first use truncated Batcher's bitonic sort on $[A_0 \ldots A_{m-1}]$ until it is sorted segment-wise per $A_i$. Next we have an efficient *CountReduce* scheme shown in Figure 2 to perform the counting. We start with array $A$, which is $[A_0 \ldots A_{m-1}]$ sorted segment-wise, and array $B$, initially containing sequential indices of elements in $A$, as shown in *kernel 1*.

In *kernel 1*, one thread is generated for each memory location of $A$. Thread with global thread index $t$ marks array $B[t]$ with $t$ if $A[t] \neq A[t+1]$. *kernel 2* "compacts" the marked positions in $B_i$ and their corresponding candidates in $A_i$ in sync to the front of each array. The "compacting" process is done by first prefetching array $A$ and $B$ to the local memory shared by a pool of threads. Each thread will then shift elements in their own assigned segments in two passes. Finally, *kernel 3* simply takes the difference between $B[t]$ and $B[t+1]$ as the frequency of candidate $A[t]$. In practice, *CountReduce* is 1.5x faster than a naive GPU parallelization.

Finally, for the ranking step, we use truncated bitonic sort to sort the key-count pairs to get the elements with top counts.

### 3.5 Memory overheads and Bucket Sharing

The full memory usage of hash tables is on the order of $O(L \cdot R \cdot Range)$, where $L$ is the number of hash tables, $R$ is the size of the reservoirs and $Range$ is the number of reservoirs per table. We ensure that if a table key is unused, then its reservoir is never materialized, so the memory overhead is less than $O(L \cdot R \cdot Range)$. Although FLASH is significantly memory efficient, we anticipate that for terabytes of data, the memory usage will still be a concern.

We present a reservoir-sharing implementation for the hash tables to reduce the memory overhead. The empirical evidence behind sharing is that most of the datasets are imbalanced in most hash tables, leaving a majority of the reservoirs containing zero or very few elements in comparison to $R$. Therefore, we allocate and initialize one shared chunk of reservoirs. Each table only keeps pointers corresponding to buckets, and when a reservoir is needed, the pointer points to a randomly selected shared reservoir. The randomly shared pool increases the space utilization significantly. As argued in Section 3.2, random sharing only hurts when two hash locations collide and are both heavy, which is an infrequent event.

We define $F \in [0, 1]$, as the ratio of the actual number of reservoirs allocated to the actual range of a hash table. As shown in figure 3, *Allocated Range = F · Actual Range*. In practice, this improves memory utilization while only marginally degrading the search quality. Figure 4 shows the accuracy trade-offs with $F$ with one of the datasets *url* described in Section 4.2. We use the same two accuracy measures described in experiment Section 4.3. Evident from the benchmarking on the *url* dataset (Figure 4), obvious quality degradation is mostly only observed for $F < 0.2$.

## 4 EXPERIMENTS AND RESULTS

### 4.1 System Details

All the experiments were performed on a single machine. The CPU benchmarks use the dual Intel(R)Xeon(R)CPUE5 - 2660v4@2.00GHz CPU with 28 physical cores and 56 parallel threads in total and a total of 512GB RAM. The hybrid CPU + GPU benchmark uses an NVIDIA Tesla P100 - PCIE with 16GB memory in addition to the CPU described above. The machine had Ubuntu 16.04 installed.

### 4.2 Datasets

We chose ultra-high dimensional datasets from different domains:

- **Webspam:** This dataset is a collection of 350k email documents using character 3-grams features. Please see [45] for details. The dimensionality of the data is over 16 million features. The average number of non-zeros is around 3700.
- **Url:** This is a collection of 2.3 million URLs with the primary aim of detecting malicious URLs (spam, phishing, exploits, etc.). The dataset has around 3.2 million features consisting of various aspects including IP address, WHOIS, Domain name, geographic location, etc. Please see [29] for details. This is a sparse dataset with around 116 non-zeros per instance.
- **Kdd12:** This is a dataset from the click-through prediction competition. It contains around 150 million instances with 54 million features. Every feature is categorical and converted to binary features according to the number of possible categories. Also, each feature vector is normalized to have unit length. This is extremely sparse data with only 11 non-zeros per instance on average.
- **Friendster:** This dataset is made from a social network available on the Stanford Large Network Collection [22]. This is a friendship graph with 65 million nodes. Every node is represented as 65 million dimensional sparse binary vector indicating a direct edge to other nodes. The average number of non-zeros is around 27. We use this dataset primarily to benchmark computations of heavy entries of large matrix multiplication outputs as shown in [35].

These datasets are from different domains and cover a variety of scales and similarity levels observed in practice. The statistics of these datasets are summarized in Table 1. The datasets are all available in *libsvm* sparse format.

All these datasets will blow up in memory if used in dense format as evident from their dimensionality. Thus, any methodology which requires centering the datasets is out of the question, as it makes the data dense. Even fast matrix multiplication, which is at the heart of most GPU based speedups is not available because sparse matrix multiplication loses all the advantages of memory coalescing.

### 4.3 Quality Metrics

We evaluate the methodologies on several different performance and accuracy measures. Our emphasis is on applications where latency is critical and therefore running time and memory are our most important concerns for a given level of accuracy.

For performance evaluations, we keep track of main memory consumption which is also the memory cost of maintaining the data structure as well as various running times. For running times, we compute the three wall clock timings: 1) Initialization, 2) Data Structure Creation (Addition) and 3) Querying. The total of these three timings indicates the total time to construct the complete k-NN graph over the full dataset from scratch.

For accuracy evaluations, we use the cosine similarity measure as the gold standard. Since this is a significantly costly operation of $O(n^2)$, we randomly selected 10000 data points and computed their neighbors and gold standard similarity with all other data points. We then calculate the following metrics, averaged over the selected 10000 data points, to report the accuracy.

**R@k**: We report the mean recall of the 1-nearest neighbor in the top-$k$ results. This is also the probability that the best neighbor appears in the top-$k$ reported elements. Following [21], we will mostly be focussing on R@100, i.e. $k = 100$, but the conclusions do not change for other values of $k$. Note that since $k$ is fixed, the precision is also determined. R@k is a popular measure to understand the recall-computation tradeoff at a given precision.

**S@k**: Also, we also report the average cosine similarity of the top-$k$ results concerning the query datapoint. This is simply the average value of the cosine similarity with the query of the top-k neighbors returned by the algorithm. Ideally, in similarity search applications such as recommendation systems, it is desirable to get few candidates with high similarity in a fraction of seconds.

### 4.4 Results

*4.4.1 Effects of Parameters on Accuracy and Performance.* For simplicity, we do not use bucket sharing, i.e., we use $F = 1$. See section 3.5 for the effect of $F$ keeping other parameters fixed. We will use *rangebits* to describe the sizes of hashtables, i.e., there will be $2^{rangebits}$ total possible buckets in each hash table. $R$ will be the reservoir size. $K$, and $L$ are the LSH parameters (Section 2.2).

Varying $R$, $K$ and $L$ controls the trade-off between the search quality, speed, and memory usage. Apparently larger values lead to more candidate pairs and increase the recall. However, they hurt the running time and memory. A Larger value of $K$ increases the number of hashes to be computed and hence hashing time for both the *Indexing* phase and the *Querying* phase. Larger values of $R$ and $L$ increases *Querying phase* timing as $k$-selection is performed on array length proportional to $R \cdot L$. We refrained from any parameter tuning. Instead, we explored the parameter space by conducting grid searches while constructing k-NN graphs of the dataset. We report the results on *url* dataset and all the corresponding numbers. Other datasets have similar conclusions.

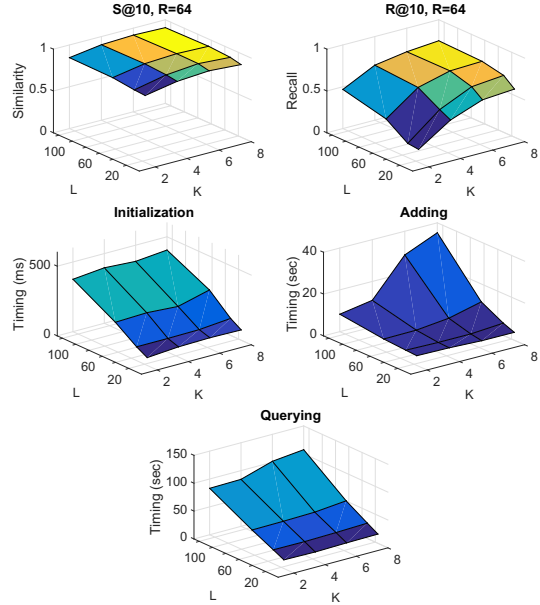

Figure 5: Effects of Varying $L$ and $K$ for *url* dataset. $R=64$.

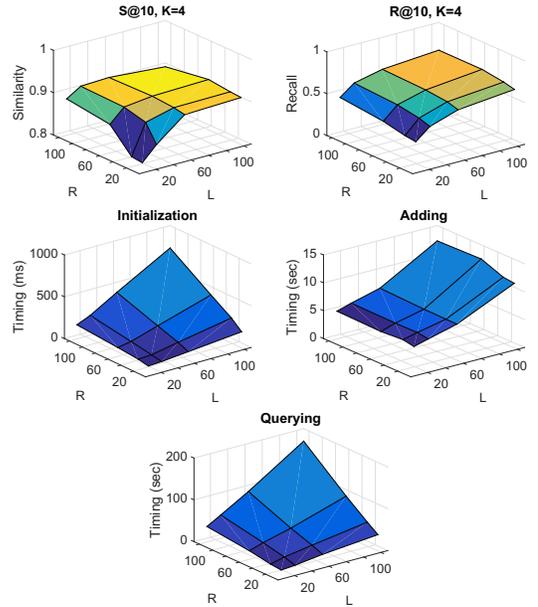

Figure 6: Effects of Varying $L$ and $R$ for *url* dataset. $K=4$.

Table 1: Dataset Information

| Dataset | Datapoints | Dimensionality (Mean Non-zero) | Mean Cosine Similarity |
|---|---|---|---|
| *url* | 2,386,130 | 3,231,961 (116) | 0.65 |
| *webspam* | 350,000 | 16,609,143 (3,728) | 0.33 |
| *kdd12* | 149,629,105 | 54,686,452 (11) | 0.15 |
| *friendster* | 65,608,366 | 65,608,366 (27.5) | Close to 0 |

Figure 5 demonstrates the effect of varying $L$ and $K$ for a fixed size of $R$. As we can see, the quality of the search is higher when $K$ and $L$ are larger. The accuracy quickly climbs to high similarity regions and then becomes flat indicating the saturation. In the saturation regions, the accuracy is very robust to the variations in the parameters. This behavior is expected from randomized algorithms. However, the runtime for all steps is almost linearly correlated, as the number of hashes to be computed and the input size for K-Selection are both proportional to $L$ and $K$.

Figure 6 demonstrates the effect of varying $R$ and $L$. In general, larger values of $R$ and $L$ lead to better search quality due to better statistical guarantees. The *Initialization* timing is linearly proportional to either $R$ or $L$ as the hash table size is linear in these parameters, hence more time is needed to initialize the memory. Since $k$-selection is performed on array length proportional to $R \cdot L$, the *Querying* timing is also nearly linearly proportional to either $R$ or $L$. The *Indexing* timing is weakly correlated to $R$ as the reservoir size is unrelated to the runtime complexity.

We usually find $K = 4$, $L = 32$, $rangebits = 15$, and $R = 32$ to give a reasonable tradeoff between accuracy and performance. We note one advantage of DOPH (minwise hashing) is that we get very accurate estimates with just 32 repetitions. Minwise hashing is known to be significantly more accurate for sparse datasets [25].

*4.4.2 Comparisons with State-of-The-Art Packages: HNSW (NMSLIB) and FAISS.* The Hierarchical Navigable Small World (HNSW) graphs method by Malkov el. al. [30] has been shown to be the best state-of-the-art ANN search method. It has been demonstrated to be the best algorithm in direct head to head comparisons with best implementations of different algorithms including FALCONN [6], *annoy* library [1], etc. Please see a very recent *ann-benchmark* by Bernhardsson [2] for detailed comparisons. NMSLIB-HNSW is known to beat the best implementation of various algorithms including LSH and trees by a significant margin on large datasets.

Thus, it suffices to evaluate FLASH against NMSLIB-HNSW thoroughly. We use the latest version, 1.6., of NMSLIB. Just like FLASH has $K$, $L$, and $R$ parameters, NMSLIB-HNSW has parameters $M$, $efConstruction$, and $efSearch$. These parameters control the number of layers and the number of search attempts, etc. for HNSW which trades running time for accuracy (See [31]). To get a fair comparison, we vary all these parameters over a fine grid and plot the recall ($R@100$) with full 100-NN graph computation time as well as the average query time on *webspam* and *url* datasets as it is fast to run several experiments on them.

NMSLIB-HNSW is a CPU based multi-threaded code. To ensure no unfair system advantage, we compare it with the CPU-only version of FLASH. To contrast the advantage of GPU based k-selection, we also show the corresponding running time over hybrid CPU-GPU version of FLASH. Figure 7 shows these tradeoffs for *webspam* and *url* dataset respectively. For better summarization, we also highlight speedups at recall level of 0.5. 0.6 and 0.7 in Table 2 along with the index size at the recall level of 0.5.

FLASH is significantly faster, both in $k$-NN construction time as well as query only time, than NMSLIB-HNSW for obtaining the same level of recall on the same system. Note the log scale on the time axis. The trends are consistent across the two datasets used. The GPU version gives another 1.5-3.5x improvement over the CPU only version, validating the superiority of our proposed k-selection.

**Higher Similarity may not mean Higher Recall:** LSH is a similarity-based retrieval method. It is known that the hardness of search based on LSH is dependent on the similarity (or distance gap) between the good neighbors and the bad neighbors. If the difference is not significant, LSH will require significant work to discriminate between them. See [18] and references therein.

In both *webspam* and *url* dataset, there is barely any difference between the similarity of the nearest-neighbor and second-best neighbor. For *webspam*, the mean similarity of the best neighbor is 0.972 while that of second best neighbor is 0.966. On *url* the best is 0.972 and the second best is 0.969. Thus, LSH cannot discriminate between them, but the recall measure R@100 is very particular about whether we get the first or the second. That is why we need to also look at other measures like S@100. However, we can quickly get the mean similarity of the best neighbor (or the S@1 measure) with FLASH to 0.941 for *webspam* and 0.955 for *url*. In practice, it will hardly make a difference if we report the 1st neighbor or 20th neighbor if their similarity with the query is substantial enough.

In section 4.4.7, we showed the aforementioned fact. For all the neighbors with similarity greater than 0.65, we get more than 90% recall in the top-20 retrieved by FLASH. It is expected that LSH will retrieve very similar points, but if the gap between the most similar and second most similar is not big, then LSH cannot discriminate between them. In practice, a high similarity is more important than the ranking. A reason why LSH can only solve the $c$-approximate near-neighbor instance which is similarity threshold based.

**Memory for Storing Index:** The most overwhelming advantage of FLASH is the reduction in index size shown in Table 2. For achieving 0.5 recall, FLASH only need 0.43GB memory and 0.2 GB with *webspam* and *url* respectively. This is 37x (16GB) and 25x (5GB) reduction on the same dataset compared to NMSLIB-HNSW. The advantages are not surprising given the simplicity of FLASH.

*4.4.3 Comparisons with NMSLIB-NSW on KDD12.* For *kdd12* dataset, which is the largest dataset in terms of $n$, NMSLIB-HNSW is significantly slower than FLASH. For NMSLIB-HNSW, we chose $efConstruction = 40$ and $M = 16$, which is one of the recommended set of parameters used for testing against FAISS on a SIFT200M dataset in the HNSW paper [31]. See Appendix E for brief details on algorithm and our usage. On kdd12, the index structured of NMSLIB-HNSW has size 122 $GiB$, which is obtained by taking the memory usage difference before and after the creation of the index. We tried attempting $efSearch = 80$ to construct the k-NN graph, but the machine goes out of memory. We used $efSearch = 20$, which completed successfully. We note that an increase in $efSearch$ increases the computational cost.

Figure 7: k-NN graph construction and per query timing comparison with HNSW on url and webspam dataset. The GPU gives another 1.5-3.5x speedup over 56 threaded CPU for the same configuration.

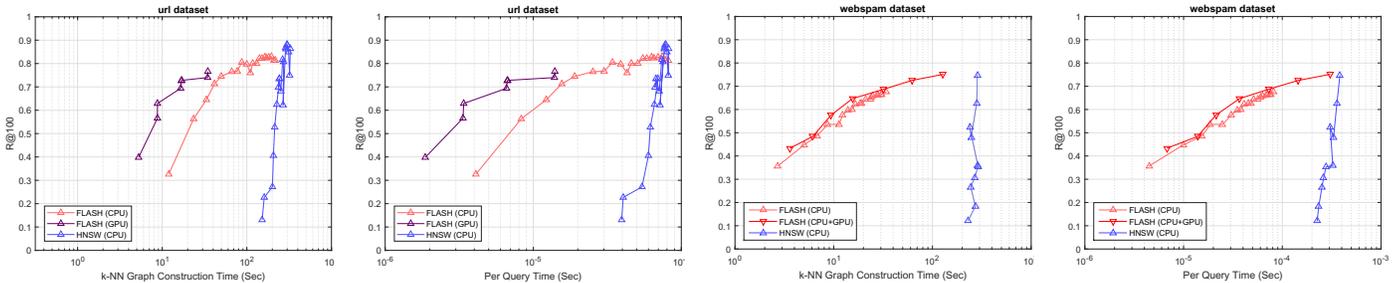

Table 2: Highlights of FLASH and HNSW comparison on url and webspam

|  | webspam | | | url | | |
| --- | --- | --- | --- | --- | --- | --- |
|  | $R@100 \approx 0.5$ | $R@100 \approx 0.6$ | $R@100 \approx 0.7$ | $R@100 \approx 0.5$ | $R@100 \approx 0.6$ | $R@100 \approx 0.7$ |
| FLASH Speedup | 27.5× | 20.2× | 9.15× | 13.6× | 9.3× | 5.8× |
|  | $R@100 \approx 0.5$ | | | $R@100 \approx 0.5$ | | |
| Index size | FLASH: 0.43 $GiB$, HNSW: 16 $GiB$ | | | FLASH: 0.2 $GiB$, HNSW: 5 $GiB$ | | |

In contrast, for FLASH (CPU only), $K = 4$, $L = 32$, $R = 64$ and hash table address size of 20-bits easily gave $R@100$ of around 0.5 which is significantly more than 0.15 obtained via the successful run of NMSLIB-HNSW. Table 3 summarizes the head-to-head comparison. FLASH only requires around 9GB of main memory for $F = 1$ and 3GB for $F = 0.3$. The index size is around 13x and 40x lower compared to 122GB. The indexing time, which is a significant bottleneck with NSMLIB-HNSW, is up to 13 times slower than FLASH and the query time of NMSLIB-HNSW is around 2x slower than FLASH on the same machine. Despite these overheads, NMSLIB-HNSW achieves poorer accuracy.

*4.4.4 GPU based FAISS Library.* The Facebook AI Similarity Search (FAISS) library is currently the best package supporting GPUs. FAISS utilizes the power of product quantization to achieve state-of-the-art ANN search with multiple GPUs [21]. However FAISS fails in the first stage as it does not provide sparse data capability, and the main memory capacity is insufficient to convert our dataset in dense format. As expected, with product quantization, every subset of dimension requires an index table, which will end up with several millions of hash tables.

To still access the performance on FAISS, we used the RCV1 dataset [23] (47k dimentions) and also generated a small sample *webspam100k* of *webspam* by only suing 100,000 sampled n-grams (random sampling to reduce dimensions) instead of 16 million and removing all samples with all zero columns. That way we can fit the dense data in GPU memory and run FAISS. We ran FAISS both in exact mode and approximate mode and compared the running time and top similarity with FLASH. The results for *webspam100k* are summarized in Table 4. We have a similar story for RCV1 data in appendix C. Clearly, FAISS is no comparison to FLASH even when dimensions are moderately high (hundred thousand).

*4.4.5 Contrast with Vanilla LSH.* We also compared the results of the standard LSH algorithm, using the same DOPH. Effectively, other than DOPH for hashing, we disable all our modification including reservoir sampling and count based k-selection. The standard algorithm uses dynamic buckets, and it computes the pairwise distance between the query and their candidates followed by ranking based on calculated distance. We found that there are some buckets which contain almost all of the data, leading to near brute force computations. To avoid this costly bruteforce, we ignore very heavy buckets from being aggregated.

We compared the performance of Vanilla LSH with FLASH on the *url* dataset using CPU only. For FLASH, we used parameters $L = 128$, $K = 4$, $R = 32$ and $rangebits = 15$. For Vanilla LSH, we used the same parameters without $R$. We obtained the following results:

**FLASH:** $S@10 = 0.901$, $S@100 = 0.856$, $R@10 = 0.640$, $R@100 = 0.783$. Runtime(Init+Indexing+Querying): $0.305 + 10.58 + 137.4 \, sec$.

**Vanilla LSH:** $S@10 = 0.838$, $S@100 = 0.781$, $R@10 = 0.186$, $R@100 = 0.187$. Runtime(Init+Indexing+Querying): $0.028 + 26.02 + 14328 \, sec$.

We can see that querying with Vanilla LSH is 100x more costly. The indexing cost is also twice as much due to the resizing overhead of the buckets. Furthermore, because heavy buckets were ignored, vanilla LSH leads to a slightly inferior accuracy. Reservoir sampling thus seems quite effective in reducing the bucket load while still randomly taking advantage of the statistics in the bucket.

Table 3: FLASH compared with HNSW on *kdd12*

| Algorithm | Indexing time | Querying time | Index Size | S@10 | R@100 |
|---|---|---|---|---|---|
| FLASH-CPU, F=1 | 4.6 *min* | 30.8 *min* | 9 *GiB* | 0.774 | 0.409 |
| FLASH-CPU, F=0.3 | 10.6 *min* | 38.6 *min* | 3 *GiB* | 0.751 | 0.280 |
| HNSW | 63 *min* | 61.9 *min* | 122 *GiB* | 0.468 | 0.156 |

Table 4: Comparison with FAISS on *webspam100k*

| Measure | FAISS Exact | FAISS Approx | FLASH |
|---|---|---|---|
| Indexing Time | 362.7 *sec* | 119.6 *sec* | 8.029 *sec* |
| Querying Time | 536.6 *sec* | 1838.2 *sec* | 1.847 *sec* |
| S@1 | 0.99 | 0.4 | 0.9 |

*4.4.6 k-NN Graph Faster than Random Projections.* A popular argument with the ultra-high dimensional dataset is first to reduce the dimensionality and then utilize any low dimensional method. In this section, we show that dimensionality reduction itself is a costly operation.

Random projection is the most efficient known algorithm for dimensionality reduction and is a prerequisite for most LSH packages. Given a data vector $v$ and $d$ random vectors $r_0 \ldots r_d$, random projection computes $v \cdot r_i \ \forall i$, forming an $d$ dimensional compressed vector. For sparse datasets, only the non-zero elements of $v$ are involved in the multiplication. We computed 100 random projections for *webspam* and *url* using the database friendly random projection [3], which is the most efficient variant of random projection for sparse datasets. We use all the 56 threads in parallel. Note, as argued before, the Fast-JL ideas based on Walsh-Hadamard [4] is not applicable as it requires making sparse data dense which will blow up the memory.

*webspam* took 671.6 sec to generate the random numbers and 426.5 sec to compute the 100 projections while *url* took 128.3 sec for generation and 65.12 sec for projection. In contrast, FLASH can compute a very reasonable full $k$-NN graph, from scratch, with $R@100 \geq 0.6$ in 20 sec for *url* and 10 sec for *webspam* on the same CPU with same number of threads. A 100-dimensional random projection is more than 42x slower than full $k$-NN graph computation with FLASH on the webspam dataset, ignoring the 671 sec for random number generation. It is worth noting that 100 dimensions are not sufficient. Usually, thousands of random projections are needed for these datasets [37].

We also contrast the time for computing random projections with the time of computing DOPH. Computing 100 DOPH hashes takes less than 1 sec for both *webspam* and *url* using 56 threads. As argued in Section 2.3, the memory and computational overheads are significant with random projections, due to large projection matrix and requirement of hundreds of passes over data. These overheads hurt the performance, especially with datasets having several millions of dimensions.

*4.4.7 Sparse Output Matrix Multiplications on a Single Machine.* In social networks, we represent the friendship relation between users as graphs. Given $d$ users, we represent each user as a $d$ dimensional sparse vector, where non-zero entries correspond to edges. By representing each user as a column, we construct matrix $A$ of dimension $d \times d$. In recommender system applications, we are interested in the finding similar users - usually friends to a similar set of users. Finding similar users requires us to compute $A^T A$, which is computationally prohibitive - with $d$ being on the million scale, there will be trillions of multiplications. Recently [35] showed a distributed algorithm based on random projection for computing heavier entries of $A^T A$ given matrix $A$.

We use Friendster data, one of the datasets used for benchmarking the computation of heavier entries of $A^T A$ in [35]. The method did not report any running time and used twitter cluster for computation. However, they require computing around 1000-2000 signed random projection as a part of their algorithm. For Friendster dataset, generating 2000 random projection, using Achlioptas method [3] takes 10220 sec on our 56 thread system, in addition to 41540 sec for random number generation. Both numbers scale linearly.

In contrast, FLASH can approximate $A^T A$ by only multiplying rows of $A^T$ with its k-nearest neighbors among columns of $A$, which essentially requires the computation of the k-NN graph. Our computation of the approximate k-NN graph of the *friendster* dataset took 26.3 minutes (or 1578 sec) and built the 20-nearest neighbor graph from scratch on a single machine. The recall of neighbors with similarity > 0.65 being 0.912 in the top 20-nearest neighbors. Thus we can find almost all heavy entries (> 0.65) of $A^T A$ quite efficiently on a single machine, which is much faster than the first step required by [35]. We use cosine similarity, but if we are interested in the inner product instead, we can resort to asymmetric minwise hashing [39] computation using DOPH.

## 5 CONCLUSION

We presented FLASH, a system for similarity search with ultra-high dimensional datasets on a single machine. FLASH uses several principled randomized techniques to overcome the computational and parallelization hurdles associated with the traditional LSH algorithm. The benefits come with strong theoretical guarantees. FLASH demonstrates the power of randomized algorithms combined with parallel processing by obtaining improvements orders of magnitude faster than state-of-the-art implementations in head-to-head comparisons. FLASH is even faster than computing random projections, which makes it naturally superior to any random projection based approach.

## 6 ACKNOWLEDGEMENTS

This work was supported by NSF IIS-1652131, NSF RI-1718478, AFOSR-YIP FA9550-18-1-0152, a gift grant from Amazon, and a GPU grant from NVIDIA. We would like to thank anonymous reviewers and Rasmus Pagh for discussions on the role of correlations in Theorem 1.

# 7 APPENDIX

## A PROOF OF THEOREM 1

The proof mimics the proof of theorem 3.4 in [17]. We give a sketch for conciseness. In the original proof, the arguments boil down to showing that the two events– E1) Good points are selected and E2) Total bad points reported are less than 3L– happens with probability higher than half. It turns out reservoir sampling, decreases the probability of retrieving candidates further, and hence the likelihood of the second event does not change. Now the LSH proof shows that (E1) holds with $1 - \frac{1}{e}$ which is more than half. We use the extra margin to incorporate failure due to reservoir sampling leading to the proof.

For good points not getting selected we need $R$ bad points to fill that particular reservoir of size $R$. Another fact that we use from the LSH proofs is that each bad point in the worst case is selected with probability $\frac{1}{n}$ in any table (part of LSH proof). Thus, the expected number of bad points in any bucket is at most 1. With DOPH, this includes the bucket containing the good point (conditional probability). Before we justify why the conditional probability is very likely to be small, we show it is sufficient.

The bucket containing the good point fails us due to reservoir sampling, with probability $(1 - \frac{1}{R})$ in the worst case. This is due to Markov inequality $Pr(\#bad \geq R) \leq \frac{1}{R}$. The modified total success probability of (E1) need to be

$$\left(1 - \frac{1}{e}\right)\left(1 - \frac{1}{R}\right) > \frac{1}{2}.$$

which gives the desired result for $R \geq 5$.

Note, here we have used the assumption that with DOPH (minwise hashing) mentioned in Section 2.3. In other words, if a bucket, where the query $q$ gets mapped, which also contains a good point $x$ (close to query $q$), does not have any higher probability of bad points $y$s (far of query) mapping in it.

This is always true if the data is independent. Even if they are not independent, our assumption is generally true for Jaccard similarity. With Jaccards similarity, as shown in Figure 8, we have the conditional probability $Pr(h(q) = h(y)|h(q) = h(x)) = \frac{a}{b} \times \frac{a}{a+y}$. This is because, we know that $h(Q) = h(X)$, therefore, the minimum of hashes of elements from both $Q$ and $X$ comes from the intersection, i.e., $a + b$. Thus, for three way collision $h(Q) = h(X) = h(Y)$, the minimum of all three should to come from $a$, instead of $a + b$, which under uniformity of hashes boils down to $\frac{a}{a+b} \times \frac{a}{a+y}$. Although it is hard to characterize the precise condition when $\frac{a}{a+b} \times \frac{a}{a+y}$ will be small given $\frac{a+b}{a+b+c+d+x+q} = S_0$ and $\frac{a+c}{a+c+q+y+b+d} = cS_0$, we can see that since $Sim(q, x) \geq S_0$ is high and $Sim(q, y) \leq cS_0$ is low ($x$ is good and $y$ is bad). $y$ and $b$ are large and $a$ is small. Thus $\frac{a}{a+b} \times \frac{a}{a+y}$ is a very small quantity, and it is likely to be smaller than $\frac{a+c}{a+c+d+b+q+y} = Pr(h(q) = h(y))$ which is our mild assumption. Thus, even with correlations we can expect our assumption to be a reasonable one.

It should be noted, that the above property is not true for projection based LSH such as signed random projection. With projections, the three-way conditional collision probability can be very hard to analyze.

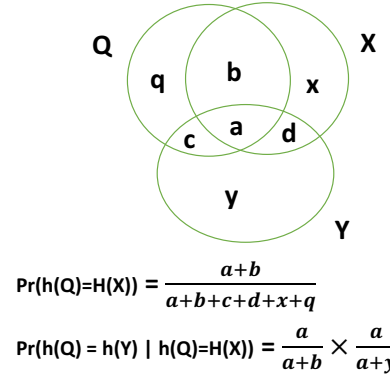

$$Pr(h(Q)=H(X)) = \frac{a+b}{a+b+c+d+x+q}$$

$$Pr(h(Q) = h(Y) \mid h(Q)=H(X)) = \frac{a}{a+b} \times \frac{a}{a+y}$$

Figure 8

Table 5: Comparison with FAISS on RCV1

| Measure | FAISS Exact | FAISS Approx | FLASH |
|---|---|---|---|
| Indexing Time | 48.3 sec | 276 sec | 1.04 sec |
| Querying Time | 1765 sec | 487.5 sec | 0.09 sec |
| S@1 | 0.6 | 0.02 | 0.2 |

## B PROOF OF THEOREM 2

It is not difficult to show that for a given query $q$ the probability of retrieving a point $x$ in any bucket is given by $1 - (1 - (Pr(h(q) = h(x)))^K)^L$. The expression is a monotonic function of $Sim(q, .)$ because the LSH is monotonic. We can further show, using elementary probability arguments, that reservoir sampling, modifies this to

$$1 - \Pi_{i=1}^{L}(1 - \alpha_i(Pr(h(q) = h(x))^K),$$

with reservoir sampling probability $\alpha_i \geq 0$, in table $i$s bucket, which is still a monotonic function of $Sim(q, .)$. The result follows from the monotonicity. Note, that every bucket has different sampling probability, but it is a fixed constant after the hash table is created.

## C FAISS VS FLASH ON RCV1 DATA

We use the popular RCV1 dataset [23] which is a benchmark for text categorization. This dataset has 47k dimensions and can comfortably fit GPU memory. The running time and accuracy comparison of FLASH vs. FAISS are summarized in Table 5. RCV1 is a low similarity dataset, so the similarity numbers are not significant, even for the exact methods. The exciting part is that for high similarity neighbors (similarity $\geq 0.85$) FLASH gets a recall of more than 80%. At the same time, it is disruptively faster, which is expected at high dimensions because product quantization methods, such as FAISS, scales poorly with dimensions.

## D MORE DISCUSSIONS ON LEVERAGING DOPH

DOPH is an efficient and practical alternative to minwise hashing. However, making DOPH to work still requires overcoming several of its limitations. DOPH, just like minwise hashing, can produce very heavy buckets where almost all (or a very large chunk of) data points will go to the same bucket in the hash table. In real-world, they are always some frequent tokens which are present in almost all the data instances. It is, therefore, possible that those tokes lead to a small hash value (with some probability) which will result in minwise hashing (or DOPH) to map all data points to the same bucket. In practice, this heavy buckets is always observed (see Section 4.4.5). These unavoidable heavy buckets will make indexing and querying a costly operation. Our work uniquely eliminates this using efficient reservoir sampling. With DOPH, we can also show theoretical guarantees of reservoir sampling, which is not true for general LSH including projection based.

Signed random projections, a popularly used LSH, can potentially avoid heavy buckets if the data is centered. Unfortunately, centering the data is not possible for ultra-high dimensions as it makes sparse data dense blowing the memory. In addition, signed random projection being 1-bit has additional problems (See section 3.3.2 for details). DOPH has large range space, and hence superior estimation via collision probability, which we have uniquely leveraged for an efficient ranking algorithm.

We have further implemented DOPH and exploited data parallelism that makes hashing time faster than data loading time. As a result of this fast implementation, we can compute a complete k-NN graph faster than computing random projections (with the same parallelism) which is the first prerequisite of most other randomized algorithms.

Effectively, our work identifies DOPH, among several other hashing schemes in the literature, to be the ideal LSH than can be leveraged to an extent where it can beat the state-of-the-art implementations.

# E  HNSW DETAILS

HNSW algorithm [30] can be seen as an extension of the probabilistic skip list structure with proximity graphs instead of the linked lists. Every inserted element will be assigned a maximum layer with an exponentially decaying probability distribution. The search starts from the top layer (the most sparse layer) by greedy traversals. After finding a local minimum on the current layer, the search will continue on the next layer, using the identified closest neighbors from the previous layer as entry points. A list of found near neighbors are kept and updated by evaluating the neighborhood of the closest previously non-evaluated element in the list. The list size starts with 1, and gradually increase as the algorithm moves on the later layers. The second phase of the algorithm begins in following layers, where the found closest neighbors on each layer are also used as candidates for the connections of the inserted element.

In our evaluation, we made use of the code contributed by the original author, under package NMSLIB-HNSW. In particular, the implementation is mainly data parallel, where parallel threads insert the elements after the very first insertion. Since this algorithm is not trivially parallelizable, locking mechanisms are employed in the implementation to prevent race conditions.